\documentclass[osajnl,preprint,showpacs,superscriptaddress,12pt]{revtex4-1} 

\usepackage{amsmath,amssymb,graphicx}

\begin{document}

\title{Nonlinear coherent state generation in the two-photon Jaynes-Cummings model}

\author{I. Ramos Prieto }

\author{B. M. Rodr\'{\i}guez-Lara}
\email{Corresponding author: bmlara@inaoep.mx}

\author{H. M. Moya-Cessa}
\affiliation{Instituto Nacional de Astrof\'{i}sica, \'{O}ptica y Electr\'{o}nica \\ Calle Luis Enrique Erro No. 1, Sta. Ma. Tonantzintla, Pue. CP 72840, M\'{e}xico}

\begin{abstract}
We show that the two-photon Jaynes-Cummings model, feasible of experimental realization in cavity or ion-trap quantum electrodynamics, can approximately produce nonlinear coherent states of the field.
We introduce these nonlinear coherent states of the field as $2m$-photon added or subtracted coherent states in terms of raising and lowering field operators, also known as London phase operators or Susskind-Glogower operators. 
\end{abstract}


\maketitle
\section{Introduction} \label{sec:S1}

Coherent states are widespread in physics \cite{Zhang1990p867}. 
In particular, coherent states defined as 
\begin{eqnarray}
\vert \alpha \rangle = e^{-\frac{\vert \alpha \vert^{2}}{2}} \sum_{j=0}^{\infty} \frac{\alpha^{j}}{\sqrt{j!}} \vert j \rangle,
\end{eqnarray}
where the states $\vert j \rangle$ are Fock or number states, are of great interest in optics as they have properties related to the classical radiation field \cite{Sudarshan1963p277,Glauber1963p2766}. 
These coherent states are eigenstates of the annihilation operator of the harmonic oscillator, 
\begin{eqnarray}
\hat{a} \vert \alpha \rangle = \alpha \vert \alpha \rangle,
\end{eqnarray}
and it is possible to generalize them as nonlinear coherent states \cite{Manko1997p528}, 
\begin{eqnarray}
f(\hat{n}) \hat{a} \vert \xi \rangle = \xi \vert \xi \rangle.
\end{eqnarray}
A class of such nonlinear coherent states are the standard single-photon-added coherent states \cite{Agarwal1991p492,Sivakumar1999p3441}, 
\begin{eqnarray}
\vert \alpha, m \rangle = \frac{1}{\sqrt{\langle  \alpha \vert \hat{a}^{m} \hat{a}^{\dagger m} \vert \alpha \rangle}} \hat{a}^{\dagger m} \vert \alpha \rangle,
\end{eqnarray} 
where the nonlinear function is given by, 
\begin{eqnarray}
f(\hat{n},m) = \frac{\hat{n} - m +1}{\hat{n} + 1},
\end{eqnarray}
and the photon number operator is $\hat{n}= \hat{a}^{\dagger} \hat{a}$.
These photon-added states can be approximately generated in the laboratory by conditional measurement in the Jaynes-Cummings dynamics \cite{Agarwal1991p492}, in a beam splitter  \cite{Dakna1998p309,Jang2014p1230}, or in spontaneous parametric down-conversion \cite{Zavatta2004p660,Zavatta2005p023820,Parigi2007p1890,Barbieri2010p063833}, to mention a few examples. 
Additional proposals to realize them in cavity or ion-trap quantum electrodynamics (QED) \cite{Dodonov1998p4087}, Kerr media \cite{RomanAcheyta2014p38}, and in quantum mechanical systems with non-linear potentials \cite{SantosSanchez2011p145307} have also been produced.
Single-photon added coherent states are known to be non-classical non-Gaussian states and, thus, useful for quantum information processing \cite{Barbieri2010p063833}.

Here, we are interested in a different class of photon-added and subtracted states based on the revival time found in the two-photon Jaynes-Cummings model \cite{Phoenix1990p116}, which is known to approximately add or subtract two photons from the initial quantized field state depending on the initial state of the qubit \cite{MoyaCessa1994p1814,MoyaCessa1999p1641}.
We propose to use the London phase operators \cite{London1926p915,London1927p193,Schleich2001}, also known as Susskind-Glogower operators \cite{Susskind1964p49}, that lower and raise the state of a quantized field, $\hat{V} \vert n \rangle = \vert n - 1 \rangle$ and $\hat{V}^{\dagger} \vert n \rangle = \vert n + 1 \rangle$, to define a general $2m$-photon added state,  
\begin{eqnarray}
\vert \psi_{+2m} \rangle= \left[ i  \hat{V}^{\dagger 2} (-1)^{\hat{n}}  \right]^{m} \vert \psi \rangle, \quad \vert \psi\rangle = \sum_{j=0}^{\infty} c_{j} \vert j \rangle,
\end{eqnarray}
that keeps the state normalized, $\langle \psi_{+m} \vert \psi_{+m} \rangle = 1 $, and adds $2m$ photons to the mean photon number of the initial state,
\begin{eqnarray}
\langle \psi_{+} \vert \hat{n} \vert \psi_{+} \rangle = \langle \psi \vert \hat{n} \vert \psi \rangle + 2m.
\end{eqnarray}
The case of $2m$-photon subtracted states can be defined in an equivalent form,
\begin{eqnarray}
\vert \psi_{-2m} \rangle= \frac{1}{\sqrt{1- \sum_{k=0}^{2m-1} \vert c_{k} \vert^{2}}} \left[ i \hat{V}^{2} (-1)^{\hat{n}} \right]^m \vert \psi \rangle, \quad \vert \psi\rangle = \sum_{j=0}^{\infty} c_{j} \vert j \rangle,
\end{eqnarray}
in order to keep the state normalized. 
This photon subtracted state will show a mean photon number that depends on the lowest Fock state components of the initial state, 
\begin{eqnarray}
\langle \psi_{-2m} \vert \hat{n} \vert \psi_{-2m} \rangle = \frac{1}{1- \sum_{k=0}^{2m-1} \vert c_{k} \vert^{2}} \left[ \langle \psi \vert \hat{n} \vert \psi \rangle -2 m+ \sum_{k=0}^{2m-1} \vert c_{k} \vert^{2} \right].
\end{eqnarray}
Note that as long as the initial state does not have the lowest $2m-1$  Fock state components, we can write
\begin{eqnarray}
 \vert \phi_{-m}\rangle= \left[i \hat{V}^{2} (-1)^{\hat{n}}\right]^{m}  \vert \phi \rangle, \quad \vert \phi\rangle = \sum_{j=2m}^{\infty} c_{j} \vert j \rangle, \label{eq:Sub}
\end{eqnarray}
and this photon subtracted state will show a mean photon number that has $2m$ less photons than the original state, 
\begin{eqnarray}
\langle \phi_{-2m} \vert \hat{n} \vert \phi_{-2m} \rangle = \langle \phi \vert \hat{n} \vert \phi \rangle - 2 m.
\end{eqnarray}
These proposed non-classical states, apart from being an experimentally feasible example of nonlinear coherent states, may be useful for quantum information processing tasks in cavity- or ion-trap-QED. 
In the following, we will show that our definition of photon added and subtracted states is experimentally feasible in cavity- and ion-trap-QED.
Then, we will present the particular case of photon added and subtracted coherent states that can be seen as nonlinear coherent states and show how to generate approximated $2m$-photon added and subtracted coherent states with the proposed experimental schemes. 

\section{A proposal for experimental realization} \label{sec:S2}

Let us consider the two-photon Jaynes-Cummings model \cite{Buck1981p132,Sukumar1984p885}, 
\begin{eqnarray}
\hat{H} = \omega \hat{a}^{\dagger} \hat{a} + \frac{\omega_{0}}{2} \hat{\sigma}_{z} + g \left( \hat{a}^{\dagger 2} \hat{\sigma}_{-} + \hat{a}^{\dagger 2} \hat{\sigma}_{-} \right),
\end{eqnarray}
describing the interaction of a quantized field and a qubit; e.g. a two-level atom interacting with the quantized field of a cavity or a two-level trapped ion interacting with the quantized motion of its center of mass.
The field is described by the frequency $\omega$ and the creation (annihilation) operators $\hat{a}^{\dagger}$ ($\hat{a}$) and the qubit by the frequency $\omega_{0}$ and the Pauli operators $\hat{\sigma}_{j}$ with $j=z,+,-$.
The interaction between the qubit and the quantized field is given by the parameter g and in both the cavity- and ion-trap-QED examples it must fulfill $g \ll \omega$.
On resonance, $2 \omega = \omega_{0}$, the evolution operator for the system is given by
\begin{eqnarray}
\hat{U}(t) = \left( \begin{array}{cc}
\cos \left[\Omega(\hat{n}) t\right] & - i  \sin \left[\Omega(\hat{n}) t\right] ~\hat{V}^{2} \\
- i \hat{V}^{\dagger 2}  \sin \left[\Omega(\hat{n}) t\right] & \cos \left[\Omega(\hat{n}-2) t\right]
\end{array} \right).
\end{eqnarray}
where we have defined the frequency $\Omega(\hat{n})= g \sqrt{(\hat{n}+2)(\hat{n}+1)}$ and used the  the lowering and raising operators defined in the introduction. 
Thus, an initial field state coupled to a qubit in the excited state, $\vert \epsilon(0) \rangle = \vert \psi, e \rangle$, will evolve as
\begin{eqnarray}
\vert \epsilon(t) \rangle =  \left( \begin{array}{cc}
\cos \left[\Omega(\hat{n}) t\right]  \\
- i  \hat{V}^{\dagger 2}  \sin \left[\Omega(\hat{n}) t\right]
\end{array} \right) \vert \epsilon(0) \rangle.
\end{eqnarray}
and for one coupled to a qubit ground state, $\vert \gamma(0) \rangle = \vert \xi, g \rangle$, its time evolution will be
\begin{eqnarray}
\vert \gamma(t) \rangle =  \left( \begin{array}{cc}
- i  \hat{V}^{2} \sin \left[\Omega(\hat{n}-2) t\right] t   \\
\cos \left[\Omega(\hat{n}-2) t\right]t
\end{array} \right) \vert \gamma(0) \rangle.
\end{eqnarray}
Note that there exists a critical Fock state $\vert j_{c} \rangle$ such that, for $j \ge j_{c}$, it is possible to approximate,
\begin{eqnarray}
\sqrt{(\hat{n}+2) (\hat{n} +1)} \vert j \rangle \approx \left( \hat{n} + \frac{3}{2} \right) \vert j \rangle,  ~ j_{c} = 3, \label{eq:Ap1}\\
\sqrt{\hat{n} (\hat{n}-1)} \vert j \rangle \approx \left( \hat{n} - \frac{1}{2} \right) \vert j \rangle, ~ j_{c}= 6 \label{eq:Ap2}.
\end{eqnarray} 
The values of $j_{c}$ in \eqref{eq:Ap1} and \eqref{eq:Ap2} guarantee a relative error between the approximation and the exact value of the order of $10^{-3}$ or less.
The origin of our definition for addition and subtraction of two-photons is that we obtain the following approximated states of the field at a time $gt = \pi$,
\begin{eqnarray}
\left\vert \epsilon \left(\frac{\pi}{g} \right) \right\rangle &\approx&   i \hat{V}^{\dagger 2} (-1)^{\hat{n}} \vert \psi, g \rangle,  \quad \vert \psi \rangle = \sum_{j=3}^{\infty} c_{j} \vert j \rangle, \\
\left\vert \gamma \left(\frac{\pi}{g}\right) \right\rangle &\approx&  i \hat{V}^{2}(-1)^{\hat{n}} \vert \psi, e \rangle, \quad \vert \psi \rangle = \sum_{j=6}^{\infty} c_{j} \vert j\rangle .
\end{eqnarray}
We require that the initial field state does not have Fock state components below the critical parameter $j_{c}$, in order to satisfy the restrictions in the approximations \eqref{eq:Ap1} and \eqref{eq:Ap2}.
Note that this allows us to fulfill the restriction in \eqref{eq:Sub}.

Thus, in a cavity-QED implementation, if the cavity field starts in the state $ \vert \psi \rangle$ and we let an atom in the excited or ground state fly through the cavity such that $gt = \pi$, then the initial cavity field will approximately end in the two-photon added or subtracted state, $\vert \psi_{+2}\rangle$  or $\vert \psi_{+2}\rangle$, as long as it did not have low Fock state components in the beginning.
The theoretically exact state of the field will be given by the reduced density matrices, 
\begin{eqnarray}
\hat{\rho}_{+2} &=& \cos\left[\Omega(\hat{n}) \pi \right] \hat{\rho}_{0} \cos\left[\Omega(\hat{n}) \pi \right] + \nonumber \\
&& + \hat{V}^{\dagger 2} \sin \left[\Omega(\hat{n})\pi\right] \hat{\rho}_{0}  \sin \left[\Omega(\hat{n})\pi\right] \hat{V}^{2}, \\
\hat{\rho}_{-2} &=& \cos\left[\Omega(\hat{n}-2) \pi \right] \hat{\rho}_{0} \cos\left[\Omega(\hat{n}-2) \pi \right] + \nonumber \\
&& + \hat{V}^{2} \sin \left[\Omega(\hat{n}-2)\pi\right] \hat{\rho}_{0}  \sin \left[\Omega(\hat{n}-2)\pi\right] \hat{V}^{\dagger 2}, 
\end{eqnarray}
where the initial state of the field, $\vert \psi \rangle$, is encoded in the density matrix $\hat{\rho}_{0}= \vert \psi \rangle \langle \psi \vert$.
This cavity field can be used as the initial field for a second atom passing and repeating the procedure $m$ times will approximately create a multiple two-photon added or subtracted state, $\vert \psi_{+2m}\rangle$ or $\vert \psi_{-2m}\rangle$ respectively.
After $m$ repetitions, the theoretically exact state of the field is 
\begin{eqnarray}
\hat{\rho}_{+2m} &=& \cos\left[\Omega(\hat{n}) \pi \right] \hat{\rho}_{2(m-1)} \cos\left[\Omega(\hat{n}) \pi \right] + \nonumber \\
&& + \hat{V}^{\dagger 2} \sin \left[\Omega(\hat{n})\pi\right] \hat{\rho}_{2(m-1)}  \sin \left[\Omega(\hat{n})\pi\right] \hat{V}^{2}, \\
\hat{\rho}_{-2m} &=& \cos\left[\Omega(\hat{n}-2) \pi \right] \hat{\rho}_{2(m-1)} \cos\left[\Omega(\hat{n}-2) \pi \right] + \nonumber \\
&& + \hat{V}^{2} \sin \left[\Omega(\hat{n}-2)\pi\right] \hat{\rho}_{2(m-1)}  \sin \left[\Omega(\hat{n}-2)\pi\right] \hat{V}^{\dagger 2}.
\end{eqnarray}
In an ion-trap-QED implementation, we can initialize the quantized center of mass motion of the atom in a suitable state that lacks low Fock state components, then start the two-photon Jaynes-Cummings dynamics with the ion in the excited or ground state and at a time such that $gt = \pi$ we can set the ion back to the excited or ground state with an auxiliary laser pulse and repeat the procedure as needed.

\section{Photon added and subtracted coherent states as nonlinear coherent states} \label{sec:S3}

It is straightforward to show that the $2m$-photon added or subtracted coherent state, $\vert \alpha_{\pm 2m} \rangle$, are eigenstates of nonlinear operators, $\hat{A}_{\pm 2m}$, with the coherent parameter as eigenvalue, 
\begin{eqnarray}
 \hat{A}_{\pm2m} \vert \alpha_{\pm 2m} \rangle = - \alpha \vert \alpha_{\pm 2m} \rangle.
\end{eqnarray}
In the case of the $2m$-photon added coherent state, 
\begin{eqnarray}
\vert \alpha_{+2m} \rangle = \left[ i  \hat{V}^{\dagger 2} (-1)^{\hat{n}}  \right]^{m} \vert \alpha \rangle,
\end{eqnarray}
the nonlinear operator is given by
\begin{eqnarray}
\hat{A}_{+2m} =  \sqrt{ \frac{\hat{n} - 2m +1}{\hat{n} + 1}} ~\hat{a}.
\end{eqnarray} 
And for the $2m$-photon subtracted coherent state, 
\begin{eqnarray}
\vert \alpha_{-2m} \rangle = \left[i \hat{V}^{2} (-1)^{\hat{n}}\right]^{m} \vert \alpha \rangle,
\end{eqnarray}
as long as the absolute value of the coherent parameter, $\vert \alpha \vert$, is large enough to guarantee that 
\begin{eqnarray}
 e^{-\vert \alpha \vert^{2}} \frac{\vert\alpha\vert^{2j}}{j!} \approx 0, ~ \mathrm{for}~ j\le 2 m,
\end{eqnarray}
in order to fulfill \eqref{eq:Sub}, the nonlinear annihilation operator is given by
\begin{eqnarray}
\hat{A}_{-2m} = \sqrt{ \frac{\hat{n}+ 2m + 1}{\hat{n} +1}} ~\hat{a}.
\end{eqnarray}

Our proposed photon adding or subtracting scheme keeps the shape of the Fock state distribution but changes the mean photon number value.
Then, it is of interest to calculate the Mandel Q parameter \cite{Mandel1995}, 
\begin{eqnarray}
Q(\psi) = \frac{ \langle \psi \vert \hat{n}^{2} \vert \psi \rangle - \langle \psi \vert \hat{n} \vert \psi \rangle^{2}}{ \langle \psi \vert \hat{n} \vert \psi \rangle} -1,
\end{eqnarray}
for our $2m$-photon added or subtracted states, 
\begin{eqnarray}
Q(\psi_{\pm 2m}) = \frac{ \langle \psi \vert \hat{n} \vert \psi \rangle }{ \langle \psi \vert \hat{n} \vert \psi \rangle \pm 2m } Q \mp \frac{2m}{\langle \psi \vert \hat{n} \vert \psi \rangle \pm 2m}.
\end{eqnarray}
In the case of our $2m$-photon added or subtracted coherent states defined above, it reduces to
\begin{eqnarray}
Q(\alpha_{\pm 2m}) =  \mp \frac{2m}{\vert \alpha \vert ^{2} \pm 2m},
\end{eqnarray}
because coherent states have a Poissonian distribution over Fock states, $Q(\alpha)=0$.
In other words, these states will always have sub- or super-Poissonian statistics, respectively. 

\begin{figure}[t]
\centerline{\includegraphics[scale=1]{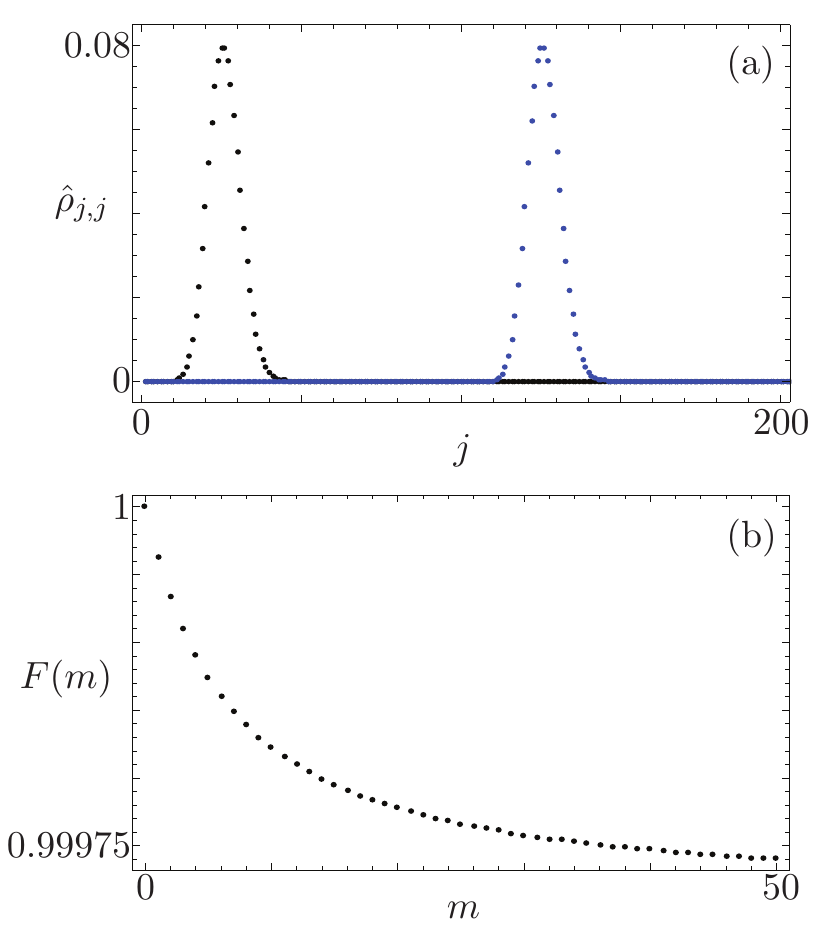}}
\caption {(Color online) (a) The Fock state distribution, $\hat{\rho}_{j,h}$, of an initial coherent state $\vert \alpha \rangle$ with (a) $\alpha=5$  (black dots) and the $100$-photon added coherent state $\vert \alpha_{+100}$ (light blue dots) obtained by applying $m=50$ times the two-photon Jaynes-Cummings procedure. (b) The fidelity, $F(m)=\mathrm{Tr}( \hat{\rho}_{+2m} \hat{\varrho}_{+2m} )$, between the exact state given by the two-photon Jaynes-Cummings procedure and the ideal $2m$-photon added coherent state. }   \label{fig:Fig1}
\end{figure}

\begin{figure}[t]
\centerline{\includegraphics[scale=1]{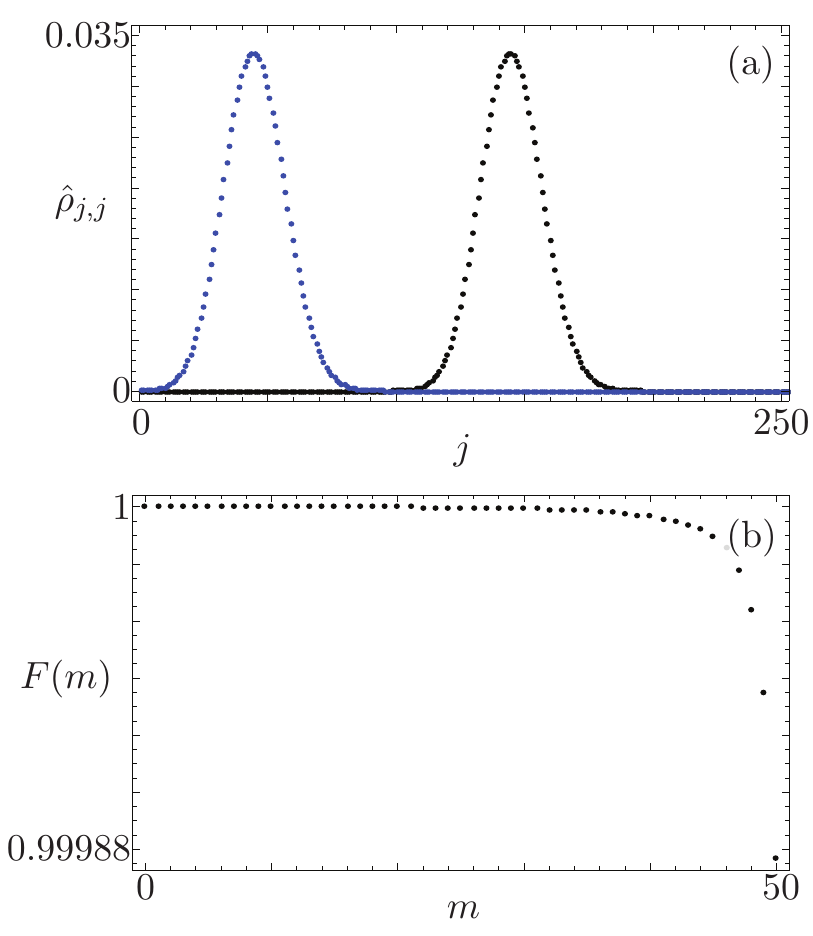}}
\caption {(Color online) (a)The Fock state distribution, $\hat{\rho}_{j,j}$, of an initial coherent state $\vert \alpha \rangle$ with (a) $\alpha=12$  (black dots) and the $100$-photon subtracted coherent state $\vert \alpha_{+100}$ (light blue dots) obtained by applying $m=50$ times the two-photon Jaynes-Cummings procedure. (b) The fidelity, $F(m)=\mathrm{Tr}( \hat{\rho}_{-2m} \hat{\varrho}_{-2m} )$, between the exact state given by the two-photon Jaynes-Cummings procedure and the ideal $2m$-photon subtracted coherent state. }   \label{fig:Fig2}
\end{figure}

It is possible to create these $2m$-photon added or subtracted in our experimental proposal by initializing the cavity field or the quantized motion of the ion in a coherent state and then implementing the adding or subtracting protocol.
Figure \ref{fig:Fig1}(a) shows the probability to find the initial (black dots) and final (light blue dots) state in the $j$th Fock state, $\hat{\rho}_{j,j}$.
The initial state is a coherent state $\vert \alpha \rangle$ with $\alpha = 5$ and the final is obtained by repeating $50$ times the two-photon adding protocol. 
The fidelity, $F(m) = \mathrm{Tr}( \hat{\rho}_{+2m} \hat{\varrho}_{+2m} )$, between the obtained field state, $\hat{\rho}_{+2m}$, and the ideal $2m$-photon added state, $\hat{\varrho}_{+2m} =  \vert \alpha_{+2m} \rangle \langle \alpha_{+2m} \vert$, is shown in Fig. \ref{fig:Fig1}(b).
Figure \ref{fig:Fig2}(a) shows the Fock state distributions for an initial coherent state with $\alpha=12$ in black dots and the final state obtained after repeating the two-photon subtracting protocol for $50$ times in light blue dots.
The corresponding fidelity, $F(m) = \mathrm{Tr}( \hat{\rho}_{-2m} \hat{\varrho}_{-2m} )$, between the obtained state, $ \hat{\rho}_{-2m}$, and the ideal $2m$-photon subtracted state $\hat{\varrho}_{-2m} =  \vert \alpha_{-2m} \rangle \langle \alpha_{-2m} \vert$ is shown in Fig. \ref{fig:Fig2}(b).

\section{Conclusions} \label{sec:S4}

We have introduced a definition of photon added and subtracted states that is experimentally feasible in cavity- and ion-trap-QED via the two-photon Jaynes-Cummings model.
These states are defined in terms of the Susskind-Glogower operators that raise or lower a Fock state and, thus, deliver a mean photon number that is just the original mean photon number plus or minus multiples of two photons.
It is important to note that while it is straightforward to define the photon-added states in this form, restrictions must be imposed on the definition of photon-subtracted states. 
Adding (subtracting) photons in this manner to (from) coherent states  makes them nonlinear coherent states with sub-(super-)Poissonian statistics that are experimentally feasible in the laboratory.

\section*{Acknowledgments}
I. Ramos Prieto acknowledges financial support from CONACYT through scholarship $\#$276331.


\end{document}